\documentclass[aps,showpacs,twocolumn,prl]{revtex4}
\usepackage{graphicx}

\usepackage{color}

\begin{document}

\newcommand{\pderiv}[2]{\frac{\partial #1}{\partial #2}}
\newcommand{\deriv}[2]{\frac{d #1}{d #2}}
\newcommand{\eq}[1]{Eq.~(\ref{#1})}  
\newcommand{\infint}{\int \limits_{-\infty}^{\infty}}

\title{Nonlinear Relativistic and Quantum Equations 
with a Common Type of Solution} 

\vskip \baselineskip

\author{F.~D. Nobre$^{1}$}
\thanks{Corresponding author: E-mail address: fdnobre@cbpf.br}

\author{M.~A. Rego-Monteiro$^{1}$}

\author{C. Tsallis$^{1,2}$}

\address{
$^{1}$Centro Brasileiro de Pesquisas F\'{\i}sicas and 
National Institute of Science and Technology for Complex Systems \\
Rua Xavier Sigaud 150 \\
22290-180 \hspace{5mm} Rio de Janeiro - RJ \hspace{5mm} Brazil \\
$^{2}$Santa Fe Institute, 1399 Hyde Park Road, Santa Fe, New Mexico 87501,
USA}

\date{\today} 

\newpage

\begin{abstract}
Generalizations of the three main equations of quantum physics, namely, 
the Schr\"odinger, Klein-Gordon, and Dirac equations, are proposed. 
Nonlinear terms, characterized by exponents
depending on an index $q$, are considered in such a way that the
standard linear equations are recovered in the limit $q \rightarrow 1$. 
Interestingly, these equations present a common, soliton-like, travelling
solution, which is  written in terms of the  
$q$-exponential function that naturally  emerges within nonextensive
statistical mechanics.  
In all cases, the well-known Einstein energy-momentum relation is preserved
for arbitrary values of $q$. 

\vskip \baselineskip


\end{abstract}
\maketitle

The applicability of linear equations in physics is usually 
restricted to idealized systems, being valid for media characterized by
specific conditions, like homogeneity, isotropy, and translational 
invariance, with particles interacting through short-range forces 
and with a dynamical behavior characterized by short-time memory.
However, many real systems -- specially the ones within the realm of 
complex systems -- do not fulfill these requirements, and  
usually exhibit complicated collective behavior associated to  
NL phenomena~\cite{scott0507}. 
Accordingly, the study of nonlinear equations (NLEs) 
has opened a new area of physics, attracting a lot of interest 
due to the possibility of describing many real situations.  
Since finding analytical solutions of NLEs can be a hard
task, particularly in the case of NL differential 
equations~\cite{polyanin}, very frequently one has to make use of 
numerical procedures and so, a considerable advance has been attained
lately. 

Among the most studied NL differential equations we have the
sine-Gordon~\cite{frenkel}, the NL 
Schr\"odinger~\cite{sulem} and Fokker-Planck ones~\cite{frankbook}. 
In these cases, the NL contributions lead to important novel insights,
relevant for modeling several physical new aspects. Two special
types of solutions have called a lot of interest, 
particularly in the sine-Gordon and NL 
Schr\"odinger equations; these are known as solitons (or solitary waves) 
and breathers~\cite{scott0507,polyanin,frenkel,sulem}. 
In both cases one has compact travelling solutions,  the first one being
characterized by a spatial part that does not deform throughout the evolution,
whereas the later presents an oscillating width as propagation occurs. 
An important property of these two solutions concerns their 
square integrability, allowing for an appropriate normalization. 
Due to the modulation of the wave function, these solutions are considered
to be relevant in diverse areas of physics, 
including nonlinear optics, superconductivity, plasma physics, and deep
water waves~\cite{scott0507}. 

In the most common NL Schr\"odinger equation, one introduces a new cubic
term in the wave  
function (see for instance \cite{sulem})
which, for some 
particular type of solution, is responsible for the modulation of the wave
function. Such addition of extra NL 
terms has been much used in the literature for constructing NLEs.
A different approach consists in transforming one or more 
linear terms into NL ones, as usually happens in NL 
Fokker-Planck equations~\cite{frankbook}. In this case, 
whereas the linear Fokker-Planck equation is associated to normal
diffusion and to the Boltzmann-Gibbs entropy, its NL counterparts are 
usually related to anomalous-diffusion 
phenomena~\cite{bouchaud90} and to generalized 
entropies~\cite{schwaemmle07}, like the nonadditive one
which yields nonextensive statistical   
mechanics~\cite{tsallis88,tsallisbook,tsallisreviewbjp09}.

This theory emerged from a generalization of the Boltzmann-Gibbs entropy,
through the introduction of 
an index $q$ ($q \rightarrow 1$ recovers the standard case).
Since then, a considerable progress was achieved, leading to generalized
functions, distributions, important equations of physics, and even new
forms of the Central Limit Theorem~\cite{CLT}.
In particular,
the $q$-Gaussian distribution, which generalizes 
the standard Gaussian, appears naturally by extremizing 
the entropy~\cite{tsallis88},  
or from the solution of the corresponding nonlinear Fokker-Planck
equation~\cite{plastino95},  and has  
been very useful for experiments in many real
systems~\cite{tsallisbook,tsallisreviewbjp09}.  
Among many others, we may
mention: 
(i) The velocities of cold atoms in dissipative optical 
lattices~\cite{douglas06};  
(ii) The velocities of particles in quasi-two dimensional dusty 
plasma~\cite{liugoreeprl08};
(iii) Single ions in radio frequency traps interacting with a classical
buffer gas~\cite{devoe}; 
(iv) The relaxation curves of RKKY spin glasses, like CuMn and
AuFe~\cite{pickup}; 
(v) Transverse momenta distributions at LHC
experiments~\cite{CMS}.

Herein we introduce NL generalizations of the three main equations 
of quantum physics, namely, 
the Schr\"odinger, Klein-Gordon, and Dirac equations.
The present proposals consist in extending linear terms into NL
ones, as done for the NL Fokker-Planck equations. 
An interesting aspect about these generalizations is that they may be
formulated easily in arbitrary dimensions, whose exact 
solutions are here presented. These solutions are all expressed in terms of
the 
$q$-exponential function $\exp_{q}(u)$ 
that naturally emerges in nonextensive 
statistics~\cite{tsallisbook,tsallisreviewbjp09}. For a pure
imaginary $iu$, one defines $\exp_{q}(iu)$ as the principal value of
\begin{equation}
\label{eq:compqexp}
\exp_{q}(iu) = \left[ 1 + (1-q)iu \, \right]^{\frac{1}{1-q}}; 
\,\exp_{1}(iu) \equiv \exp(iu). 
\end{equation}
The above function satisfies~\cite{borges98},
\begin{equation}
\label{eq:propcompqexp1}
\exp_{q}(\pm iu) = \cos_{q}(u) \pm i \sin_{q}(u)~,
\end{equation}
\vspace{-5mm}
\begin{equation}
\label{eq:propcompqexp2}
\cos_{q}(u) = \rho_{q}(u)
\cos \left\{ {1 \over q-1} {\rm arctan}[(q-1)u] \right\}~,
\end{equation}
\vspace{-5mm}
\begin{equation}
\label{eq:propcompqexp3}
\sin_{q}(u) = \rho_{q}(u)
\sin \left\{ {1 \over q-1} {\rm arctan}[(q-1)u] \right\}~,
\end{equation}
\vspace{-5mm}
\begin{equation}
\label{eq:propcompqexp4}
\rho_{q}(u) = \left[1+(1-q)^{2}u^{2} \right]^{1/[2(1-q)]}~,
\end{equation}
\vspace{-5mm}
\begin{equation}
\label{eq:propcompqexp5}
\exp_{q}(iu)\exp_{q}(-iu) = \cos_{q}^{2}(u) + \sin_{q}^{2}(u) = 
\rho_{q}^{2}(u)~. 
\end{equation}
Notice that $\exp_{q} \left[ i(u_{1} + u_{2}) \right]
\ne \exp_{q} (i u_{1})  \exp_{q} (i u_{2})$ for $q\ne 1$
\cite{tsallisbook}.  
As a consequence of Eqs.~(\ref{eq:propcompqexp1}-\ref{eq:propcompqexp5}),
the $q$-exponential  
of a pure imaginary presents an oscillatory behavior with a varying
amplitude $\rho_{q}(u)$ that decreases (increases) for $1<q<3$ \ ($q<1$).
By integrating~\eq{eq:propcompqexp5} from $-\infty$ to $+\infty$, we verify
the physically important property of square integrability for  
$1<q<3$, whereas such integral diverges in both limits 
$q \rightarrow 1$ and $q \rightarrow 3$ and for $q<1$~\cite{max}. 
More precisely, $\exp_{q}(iu)$  
is modulated by the $q$-Gaussian in~\eq{eq:propcompqexp4}. 
This is the type of solution that we focus on throughout the
present paper.  

Let us first consider the simple one-dimensional linear wave equation 
\begin{equation}
\label{eq:1dlwaveeq}
{\partial ^{2} \Phi(x,t) \over \partial x^{2}} = 
{1 \over c^{2}}
{\partial ^{2} \Phi(x,t) \over \partial t^{2}}~,   
\end{equation}
for which any function of the type $\Phi(kx-\omega t)$, twice differentiable,
is a solution. In particular, one may have a {\it $q$-plane wave}~\cite{max},   
\begin{equation}
\label{eq:qsolwaveeq}
\Phi(x,t) = \Phi_{0} \, \exp_{q} \left[ i (kx-\omega t) \right]~; \quad 
[\Phi_{0} \equiv \Phi(0,0)]~, 
\end{equation}
as a solution of the equation above, 
provided that $\omega = c \, k$. 
Let us remind that the above $q$-plane wave cannot be decomposed, for
$q\ne1$, into spatial and temporal factors, as it 
happens in many solutions of NLEs~\cite{polyanin}. Moreover,  
the amplitude of the $q$-plane wave for $1<q<3$ decreases
when its argument $(kx-\omega t)$ increases. 
Due to this property, this new type of solution may well be applicable to
various  nonlinear physical phenomena. 
Let us stress that, in the particular situation where $x=ct$, one has 
$\Phi(x,t) = \Phi_{0} \ (\forall t)$, consequently 
the $q$-plane wave behaves like a soliton propagating with a velocity 
$c= \omega/k$. This enables the approach of nonlinear
excitations which do not deform in time and should be relevant, e.g., 
in nonlinear optics and plasma physics. 

Now, since~\eq{eq:1dlwaveeq} is a linear equation, 
the following linear combination,  
\begin{equation}
\label{eq:linqsolwaveeq}
\Phi(x,t) = \sum_{j} \Phi_{0j} \, \exp_{q_{j}} \left[ i (kx-\omega t) \right]~, 
\end{equation}
is also a solution {\it with the same dispersion relation}. 
A given physical system may be characterized by a single value of $q$. If 
$1<q<3$, we have $\Phi(\pm \infty,t)=0 \ (\forall t)$; by specifying the
rate of decay of the  
$q$-plane wave amplitude (which should be a characteristic of a given 
physical system), i.e., 
its modulation, \eq{eq:propcompqexp4}, we may determine the 
appropriate value of $q$. 

Another important aspect of the $q$-plane 
wave concerns its immediate extension to $d$ dimensions, 
\begin{equation}
\label{eq:3dqsolwaveeq}
\Phi(\vec{x},t) = \Phi_{0} \, \exp_{q} \left[ i (\vec{k} \cdot \vec{x}
-\omega t) \right]~, 
\end{equation}
which, due to the well-known properties of the scalar product 
$\vec{k} \cdot \vec{x}= \sum_{n=1}^{d} k_{n}x_{n}$, exhibits invariance
under rotations. 
If we take into account that $d\exp_q(z)/dz=[\exp_q(z)]^q$ and
$d^2\exp_q(z)/dz^2=q[\exp_q(z)]^{2q-1}$ we obtain, for the $d$-dimensional
Laplacian, 
\begin{equation}
\label{eq:nabla2qexp}
\nabla^{2} \Phi(\vec{x},t) = 
- q \left( \sum_{n=1}^{d} k_{n}^{2} \right) \Phi_{0} 
\{ \exp_{q} [  i (\vec{k} \cdot \vec{x}
-\omega t)] \}^{2q-1}~.  
\end{equation}
>From the results above one sees that the 
$d$-dimensional $q$-plane wave of~\eq{eq:3dqsolwaveeq} satisfies the
linear wave equation, 
\begin{equation}
\label{eq:3dlwaveeq}
\nabla^{2} \Phi(\vec{x},t) = 
{1 \over c^{2}}
{\partial ^{2} \Phi(\vec{x},t) \over \partial t^{2}} \quad \Rightarrow 
\quad \omega=ck ~,   
\end{equation}
with $k = \sqrt{\sum_{n=1}^{d} k_{n}^{2}}$ , showing that the resulting
dispersion relation is $q$-invariant.  

We will now focus on NLEs for which the above
$d$-dimensional $q$-plane wave is an exact solution. It should be
emphasized at 
this point that many NLEs in the literature are usually formulated in 
one dimension and that their extension to  $d$ dimensions  
may not be always an easy task.
Accordingly, let us introduce the following $d$-dimensional 
NL generalization of
the Schr\"odinger equation for a particle of mass $m$, 
\begin{equation}
\label{eq:schreq}
i \hbar {\partial \over \partial t} 
\left[ \frac{\Phi(\vec{x},t)}{\Phi_{0}} \right] 
= - {1 \over 2-q} \ \frac{\hbar ^{2}}{2m}
\nabla^{2} \left[ \frac{\Phi(\vec{x},t)}{\Phi_{0}} \right]^{2-q}~.  
\end{equation}
We notice that the scaling of the wave function by 
$\Phi_{0}$ guarantees the correct physical dimensionalities for all terms.
This scaling becomes 
irrelevant only for linear equations [e.g., 
in the particular case $q=1$ of~\eq{eq:schreq}]. 
Consistently, the energy and momentum operators are generalized as 
$\hat{E} = i \hbar D_{t}$ and $\hat{p}_{n} = -i \hbar D_{x_{n}}$
respectively, where $D_{u} f(u) \equiv [f(u)]^{1-q} d f(u)/du$.
These operators, when acting on the $q$-exponential
$\exp_{q} \left[ i (\vec{k} \cdot \vec{x}-\omega t) \right]$, yield 
the energy $E = \hbar \omega$ and momentum $\vec{p} = \hbar \vec{k}$.
Now, if ones considers the $q$-plane wave solution  
of~\eq{eq:3dqsolwaveeq} by using 
$\vec{k} \rightarrow \vec{p}/\hbar$ and  
$\omega \rightarrow E/\hbar$, one verifies that this new form is a solution 
of the equation above, with $E=p^{2}/2m$, {\it for all values of $q$}. 
The NL Schr\"odinger equation of~\eq{eq:schreq} shows the
same structure of the NL Fokker-Planck 
equation of Refs.~\cite{plastino95} in the absence of an 
external potential, which appears in 
nonextensive statistical mechanics~\cite{tsallisbook,tsallis88}.
Essentially, it represents the analogue of the porous-medium 
equation~\cite{vazquezbook}, very common in 
anomalous-diffusion phenomena~\cite{bouchaud90}, but with an imaginary
time. One well-known solution of the porous-medium 
equation is due to 
Barenblatt~\cite{barenblatt}. This solution presents a structure very 
similar to a $q$-Gaussian, written in terms of a power law
like the one in~\eq{eq:qsolwaveeq}.

However, \eq{eq:schreq} differs from previous 
formulations~\cite{sulem}, where one
adds a new NL term (in most cases, a cubic nonlinearity in the wave
function) to the two existing linear terms. This extra NL term 
is responsible for the modulation of the wave function, which has made such
equations relevant for  
many physical phenomena since some decades. Particularly, the breather-like
solutions 
are characterized by a 
time dependence of the linear type, i.e., ${\rm e}^{i \omega t}$, 
such that their nonlinearities are essentially manifested 
through their spatial dependences. In its discrete formulation, this equation
leads to discrete breathers, much studied in the
literature due to their potential applications in lattice dynamics of
solids, coupled arrays of Josephson junctions, and localization of waves 
in random lattices,
among others~\cite{flach,kopidakis0108}. The main differences
between~\eq{eq:schreq} and the existing forms of NL Schr\"odinger 
equations in the literature are: 
(i) Instead of adding an extra term in which the nonlinearity is
introduced, we modify the spatial second-derivative term; 
(ii) The equation, together with the proposed solution, are  
easily extended from one to $d$ dimensions;  
(iii) The corresponding solution of~\eq{eq:schreq} manifests nonlinearity
in both space and time, through a modulation in these two variables,  
which keeps the norm finite for all $(\vec{x},t)$; 
(iv) The well-known energy spectrum $E=p^{2}/2m$ is preserved 
for all $q$. 
Therefore \eq{eq:schreq}, together
with the simple solution in~\eq{eq:3dqsolwaveeq}, emerge 
as possible descriptions of physical 
phenomena, like the propagation of nonlinear pulses in optical fibers for
carrying information (``bits'').

Another category of NL equations studied in the literature concerns the
class of NL Klein-Gordon 
equations (see for 
instance~\cite{polyanin,frenkel,kopidakis0108,hioe03,hioe04}). 
As in the previous case, in most NL Klein-Gordon equations 
the second-derivative terms are left unchanged and the
nonlinearity is introduced by means of extra terms containing 
powers of the Klein-Gordon field~\cite{polyanin}, e.g.,  
cubic~\cite{hioe03,hioe04} or quartic~\cite{kopidakis0108} terms.
If one extends this nonlinearity to a general 
functional of the field, then the sine-Gordon
equation~\cite{frenkel}, which contains a term expressed as 
a sine of the field, may be also included in this 
category~\cite{scott0507,flach}. 
Several types of solutions have been proposed in the
literature for such NL Klein-Gordon equations, particularly those 
of the Barenblatt kind~\cite{barenblatt} and breather-like 
ones~\cite{flach,kopidakis0108}. 
In other cases the solutions are characterized by an
amplitude that may decay in both space and time, multiplied by a
periodically oscillating part in the form
${\rm e}^{i (Ax+Bt)}$ ($A$ and $B$ constants)~\cite{hioe03,hioe04}.
The solution presented in~\cite{hioe03,hioe04} is qualitatively similar to
the one-dimensional $q$-plane wave. However, it may not be trivially
extended to $d$-dimensions.  

Herein we propose a new NL Klein-Gordon equation in $d$ dimensions, namely 
\begin{equation}
\label{eq:kgordoneq}
\nabla^{2} \Phi(\vec{x},t) = 
{1 \over c^{2}}
{\partial ^{2} \Phi(\vec{x},t) \over \partial t^{2}} + 
q \, {m^{2}c^{2} \over \hbar^{2}} \, \Phi(\vec{x},t) \left[
{\Phi(\vec{x},t) \over \Phi_{0}} \right]^{2(q-1)}~.  
\end{equation}
One may verify
easily that the same $q$-plane wave used for the NL Schr\"odinger 
equation is a solution 
of~\eq{eq:kgordoneq}, {\it preserving for all $q$ the Einstein relation}  
\begin{equation}
\label{eq:einsteinrel}
E^{2} = p^{2}c^{2} + m^{2}c^{4}~. 
\end{equation}
It should be mentioned that, to our knowledge, the nonlinear term 
in~\eq{eq:kgordoneq} is new and different from those 
of previous formulations, even in the particular cases 
$q=2$ and $q=5/2$, where one gets cubic and quartic terms in the field,
respectively. In previous works, the nonlinear term is 
constructed by multiplying the wave function by a power of its modulus,
leading to different types of solutions and {\it different energy spectra}.
Another important aspect of~\eq{eq:kgordoneq} concerns its Lorentz
invariance: since this property is directly related 
to the first two terms (containing derivatives)~\cite{liboff}, 
which have not been
changed herein, then~\eq{eq:kgordoneq} remains invariant under 
Lorentz transformation. Some previous NL Klein-Gordon equations 
are not invariant under usual 
Lorentz transformation~\cite{hioe03,hioe04} and require its 
generalization~\cite{hioe04}.  

Along the same line we introduce a NL form for the
Dirac equation. The present proposal represents, to the best 
of our knowledge, a new one (see 
Ref.~\cite{fushchich} for a previous, different formulation).
In this case, we will restrict ourselves to $d=3$
spatial dimensions; let us then introduce the following generalized 
Dirac equation,
\begin{equation}
\label{eq:diraceq}
i \hbar \ {\partial \Phi(\vec{x},t) \over \partial t}
+ i \hbar c (\vec{\alpha} \cdot \vec{\nabla}) \Phi(\vec{x},t)
= \beta mc^{2} A^{(q)}(\vec{x},t) \ \Phi(\vec{x},t)~, 
\end{equation}
where $\alpha_{x},\alpha_{y},\alpha_{z}$ (written in terms of the Pauli
spin matrices) and $\beta$ (written in terms of the $2 \times 2$ identity
matrix $I$) 
are the standard $4 \times 4$ matrices~\cite{liboff}. 
The new, $q$-dependent, term is given by the $4 \times 4$ diagonal matrix
$A^{(q)}_{ij}(\vec{x},t) = \delta_{ij}[\Phi_{j}(\vec{x},t)/a_{j}]^{q-1}$,
where $\{a_{j}\}$ are complex constants
($A^{(1)}_{ij}(\vec{x},t)=\delta_{ij}$).  
The solution of~\eq{eq:diraceq} we focus on is the following
four-component column matrix 
\begin{equation}
\label{eq:quadsol}
\Phi(\vec{x},t) \equiv 
\left( 
\begin{array}{c}
\Phi_{1}(\vec{x},t) \\
\Phi_{2}(\vec{x},t) \\
\Phi_{3}(\vec{x},t) \\
\Phi_{4}(\vec{x},t)
\end{array}
\right) 
= 
\left( 
\begin{array}{c}
a_{1} \\
a_{2} \\
a_{3} \\
a_{4}
\end{array}
\right) 
\, \exp_{q} \left[ {i \over \hbar} 
(\vec{p} \cdot \vec{x} - Et) \right]~.
\end{equation}
Substituting this four-component vector 
into~\eq{eq:diraceq}, we get, for the 
coefficients $\{ a_{j} \}$, {\it precisely the same set of four algebraic
equations  corresponding to the linear case} (see page 803, Eq. (15.45b) of
\cite{liboff}). These equations have, for all $q$, a nontrivial solution 
only if the Einstein energy-momentum relation (\eq{eq:einsteinrel})
is satisfied. 
The proposal in~\eq{eq:diraceq} differs from that in~\cite{fushchich} in the same 
sense of the herein proposed NL
Schr\"odinger and Klein-Gordon equations, i.e., the 
nonlinearity is introduced by generalizing an existing term, rather than 
adding an extra NL term. As  
a consequence, our solutions are given by the above-mentioned 
$q$-exponential, thus preserving relation (\ref{eq:einsteinrel}), in contrast 
to those in Ref.~\cite{fushchich}, which are typically 
written in terms of the standard exponential.
The positive and negative parts of the present 
energy spectrum are naturally expected to be respectively
associated to a particle and its corresponding 
antiparticle. 
Contrary to the standard plane waves (case $q=1$), one has
$q$-plane wave solutions, which are  
square-integrable, i.e., with a finite norm for $1<q<3$, representing 
localized particles.

To conclude, we have shown that the $q$-plane waves, which consist of a
generalization -- within the 
framework of nonextensive statistical mechanics --
of the standard plane waves,  
represent a very general type of solutions for some physical nonlinear
equations. Besides the $d$-dimensional  
linear wave equation, it appears also as a solution of important nonlinear
generalizations of 
the main equations of quantum physics, namely, nonlinear 
$d$-dimensional
Schr\"odinger and Klein-Gordon equations, and a three-dimensional 
Dirac equation (its generalization to $d$-dimensions should in principle be
straightforward; see~\cite{peskin} for the linear case).  
In contrast to other generalizations known in the literature, 
where nonlinear terms are added to the usual linear ones -- and whose
extensions 
from one- to $d$-dimensional space are in many cases
not particularly straightforward -- the present 
proposals may be easily formulated in $d$ dimensions.  They 
consist in fact in modifying existing terms by
introducing powers characterized by an index $q$. The
standard linear equations are recovered in the limit $q \rightarrow 1$. 
Furthermore, the equations and solutions presented here preserve, in all
cases, the usual energy-momentum relations.
Due to its simplicity and properties described herein, this type of
solution represents a good candidate for describing several nonlinear
physical phenomena characterized by oscillatory motion 
with modulation in both space and time,   
like those appearing in superconductivity, 
plasma physics, nonlinear optics, and lattice dynamics of solids.
Naturally, at the present early stage, specific applications of this theory 
are still elusive. 

We thank Prof. E.~M.~F. Curado for fruitful conversations. 
The partial financial supports from
CNPq and FAPERJ (Brazilian agencies) are acknowledged. 


\end{document}